\documentclass[a4paper,twocolumn,11pt,accepted=2022-03-03]{quantumarticle}
\pdfoutput=1
\usepackage{graphicx,epstopdf}
\usepackage{bm,mathtools}
\usepackage{amsmath,amssymb}
\usepackage{mathrsfs}
\usepackage{color}
\usepackage{hyperref}
\usepackage[numbers,sort&compress]{natbib}
\usepackage{CJK}
\usepackage{lipsum}

\begin{document}
%\begin{CJK*}{UTF8}{gbsn}

\title{Photon-photon interactions in Rydberg-atom arrays}

%\author{Lida Zhang~(\CJKfamily{gbsn}张理达)}
\author{Lida Zhang}
\affiliation{Center for Complex Quantum Systems, Department of Physics and Astronomy, Aarhus University, DK-8000 Aarhus C, Denmark}

\author{Valentin Walther}
\affiliation{Center for Complex Quantum Systems, Department of Physics and Astronomy, Aarhus University, DK-8000 Aarhus C, Denmark}
\affiliation{ITAMP, Harvard-Smithsonian Center for Astrophysics, Cambridge, Massachusetts 02138, USA}

\author{Klaus M{\o}lmer}
\affiliation{Center for Complex Quantum Systems, Department of Physics and Astronomy, Aarhus University, DK-8000 Aarhus C, Denmark}

\author{Thomas Pohl}
\affiliation{Center for Complex Quantum Systems, Department of Physics and Astronomy, Aarhus University, DK-8000 Aarhus C, Denmark}

%\date{\today}
\maketitle
%\end{CJK*}
\begin{abstract}
We investigate the interaction of weak light fields with two-dimensional lattices of atoms with high lying atomic Rydberg states. This system features different interactions that act on disparate length scales, from zero-range defect scattering of atomic excitations and finite-range dipolar exchange processes to long-range Rydberg-state interactions, which span the entire array and can block multiple Rydberg excitations. Analyzing their interplay, we identify conditions that yield a nonlinear quantum mirror which coherently splits incident fields into correlated photon-pairs in a single transverse mode, while transmitting single photons unaffected. In particular, we find strong anti-bunching of the transmitted light with equal-time pair correlations that decrease exponentially with an increasing range of the Rydberg blockade.
Such strong photon-photon interactions in the absence of photon losses open up promising avenues for the generation and manipulation of quantum light, and the exploration of many-body phenomena with interacting photons.
\end{abstract}

Photons typically cross each other unimpeded. Yet, the scientific and technological prospects of effective photon interactions \cite{Chang2014NPhoton} have motivated substantial research efforts into nonlinear optical processes at the ultimate quantum level. Here, optical resonators \cite{Birnbaum2005Nature,Volz2014,Reiserer2015,Welte2018,Shea2013} and nano-scale photonic structures \cite{Goban2012PRL,Thompson2013Science,Tiecke2014Nature,Petersen2014Science,Lodahl2015RMP,Chang2018RMP,Noaman2018,Yu2019,Prasad2020} have made it possible to couple photons to single saturable emitters, and strong interactions between highly excited atoms have been used to realize large optical nonlinearities in atomic ensembles \cite{Pritchard2010,Peyronel2012Nature,Thompson2017Nature,Mandoki2017,murray2016a,Firstenberg_2016}. The use of many-particle systems to reach a strong collective light-matter coupling present an attractive approach, and the exploitation of Rydberg-state interactions in atomic gases has indeed enabled recent breakthroughs that, for example, demonstrated single-photon switching \cite{Baur2014PRL,Gorniaczyk2014PRL,Tiarks2014PRL} and photonic quantum gates \cite{Tiarks2019NPhys}. Yet, photon losses that are intrinsic to such ensemble approaches \cite{Gorshkov2013PRL,Murray2018PRL} limit the performance of these applications \cite{Baur2014PRL,Murray_2016b} and severely challenge the exploration of correlated quantum states \cite{Otterbach2013} beyond the few-photon regime \cite{Zeuthen2017,Bienias2020}. At the same time, studies of ordered arrangements, instead of random atomic ensembles, have revealed a number of exciting linear optical properties \cite{Facchinetti2016PRL,Perczel2017,Bettles2017,Guimond2019PRL,Ballantine2020} that arise from the many-body nature of light-matter interactions in these systems. In particular, the cooperative response of two-dimensional arrays facilitates strong photon coupling at greatly reduced losses \cite{Bettles2016PRL,Shahmoon2017PRL}, as recently demonstrated experimentally with ultracold atoms in optical lattices \cite{Rui2020Nature}.

\begin{figure}[thb!]
 \centering
 \includegraphics[width=0.45\textwidth]{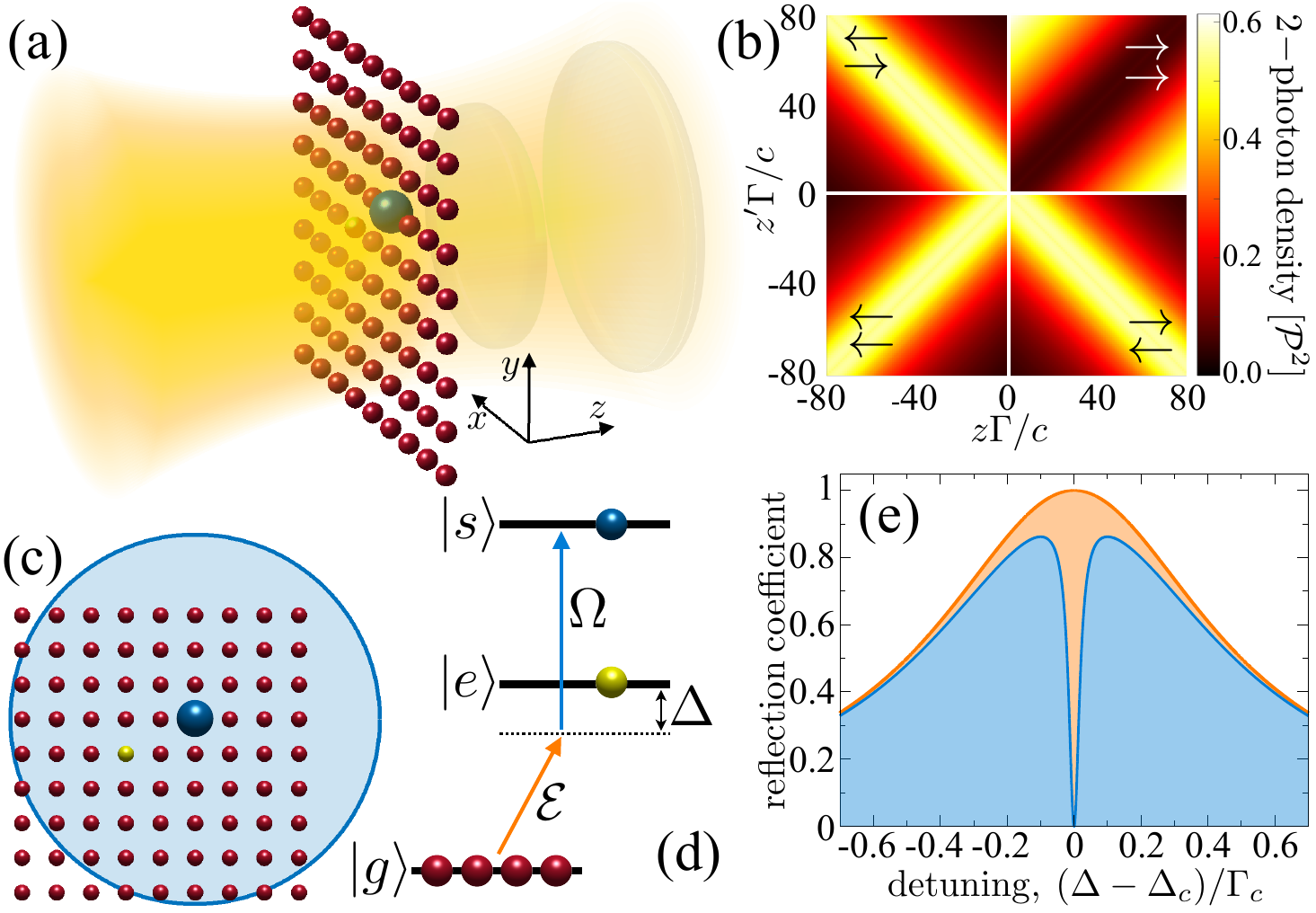}
 \caption{(a) A regular array of atoms interacts with weak coherent light and can induce nonclassical correlations in the transmitted probe field. (d) Its amplitude, $\mathcal{E}$, couples the ground states, $|g\rangle$, of the atoms to an excited state, $|e\rangle$, which is coupled to a high-lying Rydberg state $|s\rangle$ by an additional control field with a Rabi frequency $\Omega$. Atomic interactions, induced by the driven dipoles of the lower transition, lead to a collective energy shift, $\Delta_c$, and collective photon emission of the excited array with a rate $\Gamma_c$. For $\Omega=0$, this can result in near-perfect reflection [red line in (e)] when the probe-field detuning $\Delta$ matches $\Delta_c$, while EIT of the three-level system yields perfect transmission on two-photon resonance for a finite control-field coupling ($\Omega\neq0$) [blue line in (e)]. (c) The van der Waals    interaction between Rydberg atoms can be strong enough to inhibit the excitation of more than a single Rydberg state within a blockade area (blue circle) that may cover the entire array. In combination with EIT and the collective photon reflection of the array, this provides a nonlinear mechanism for strong coherent photon interactions that can generate highly non-classical states of light. This is shown by the strong pair correlations of transmitted ($\rightarrow$) and reflected ($\leftarrow$) photons in panel (b).}
 \label{fig1}
\end{figure}

In this work, we investigate the effects of strong Rydberg-state interactions in regular atomic arrays [Fig.\ref{fig1}(a)], and analyse their \emph{nonlinear cooperative} response. Rydberg-state interactions can be used to couple a single atom to adjacent lattices  \cite{Grankin2018,Bekenstein2020}, and here we show that atomic interactions within Rydberg arrays can generate strong photon-photon interactions at greatly suppressed losses. This quantum optical nonlinearity emerges from the interplay of various atomic interactions and a narrow transmission feature [Fig.\ref{fig1}(e)] that arises from three-level photon coupling [Fig.\ref{fig1}(d)] under conditions of electromagnetically induced transparency (EIT), and can produce highly correlated states of light [see Fig.\ref{fig1}(b)]. Compared to two-level systems, where few-photon nonlinearities can also arise in small arrays with very small distances, $\lesssim100$nm \cite{Cidrim2020,Williamson2020}, or be induced via Rydberg-dressing of low-lying states \cite{Moreno2021,henkel2010}, Rydberg-EIT in the present lattice-setting provides strong photon coupling and facilitates large nonlinearities under conditions of present experiments \cite{Rui2020Nature,Zeiher2015}.

We consider a two dimensional regular array of closely spaced atoms at positions ${\bf r}_j$, as illustrated in Fig.\ref{fig1}(a). A weak probe field with an amplitude $\mathcal{E}$ drives the transition between the ground state $|g\rangle$ and an intermediate state $|e\rangle$ at a frequency detuning $\Delta$, while a high-lying Rydberg state $|s\rangle$ is excited by an additional control field with a Rabi frequency $\Omega$ [Fig.\ref{fig1}(d)]. Throughout this article, we consider driving on two-photon resonance, leaving only a finite single-photon detuning $\Delta$. 
The combined action of both light fields leads to two distinct types of atomic interactions that    act on vastly different length scales. 

First, the Rydberg atoms feature van der Waals interactions that can be sufficiently strong to block the excitation of multiple Rydberg $|s\rangle$-states within distances of several micrometers \cite{Jaksch2000PRL,Lukin2001PRL}. This Rydberg blockade has been explored for a range of applications \cite{Saffman2010RMP,Adams2019JPB}. In particular, it has already been demonstrated in dense atomic lattices \cite{Zeiher2015}, with an excitation blockade over distances of more than $\sim10$ sites. We focus here on configurations in which the entire atomic array is covered by the blockade radius, and quantum states with more than a single Rydberg excitation are blocked by the strong atomic interaction [see Fig.\ref{fig1}(c)]. This full blockade regime is described  efficiently within a truncated Hilbert space that contains only  many-body states of the array with not more than a single Rydberg-excitations, while retaining states with multiple low-lying $|e\rangle$-excitations.

Second, a small lattice constant $a\sim\lambda$, on the order of the $|g\rangle-|e\rangle$ transition wavelength $\lambda$, entails strong dipole-dipole interactions that arise from near-resonant photon exchange on the probe transition \cite{James93,Dung2002PRA,Garcia2017PRA}, which leads to coherent exchange of atomic $|e\rangle$-excitations across the atomic array. This results in a collective optical response that can greatly suppress photon scattering and generate near-perfect coherent coupling to the single transverse mode of the incident field. One can find superradiant as well as subradiant states of a single de-localized $|e\rangle$-excitation, whose collective emission rate $\Gamma_c$ is respectively enhanced or suppressed relative to the single-atom decay rate $\Gamma$~\cite{Zoubi2011PRA,Sutherland2016PRA,Facchinetti2016PRL,Bettles2015PRA,Guimond2019PRL,Zhang2019PRL,Orioli2019PRL}. For large extended arrays, the resulting collective level shift \cite{Glicenstein2020}, $\Delta_c$ of the $|g\rangle-|e\rangle$ transition marks the spectral position of reflection resonances at which an incoming photon is reflected perfectly~\cite{Bettles2016PRL,Shahmoon2017PRL}, without scattering into other transverse modes.

Moreover, the control-field coupling to the Rydberg-state permits to control  the optical response on the $|g\rangle - |e\rangle$ transition. In particular, on two-photon resonance, the three-level system features a dark eigenstate that does not contain the intermediate state $|e\rangle$~\cite{Fleischhauer2000PRL,ARIMONDO1996257}, and therefore enables lossless transmission of the incident light due to EIT \cite{Fleischhauer2005RMP}. As illustrated in Fig.~\ref{fig1}(e), EIT is restricted to a narrow transparency window in the reflection spectrum of the array. Its width, $\Omega^2/\Gamma_c$, permits to control reflectivity by tuning the intensity of the classical control field \cite{Manzoni_2018,Bekenstein2020}.

\begin{figure}[t!]
 \centering
 \includegraphics[width=0.48\textwidth]{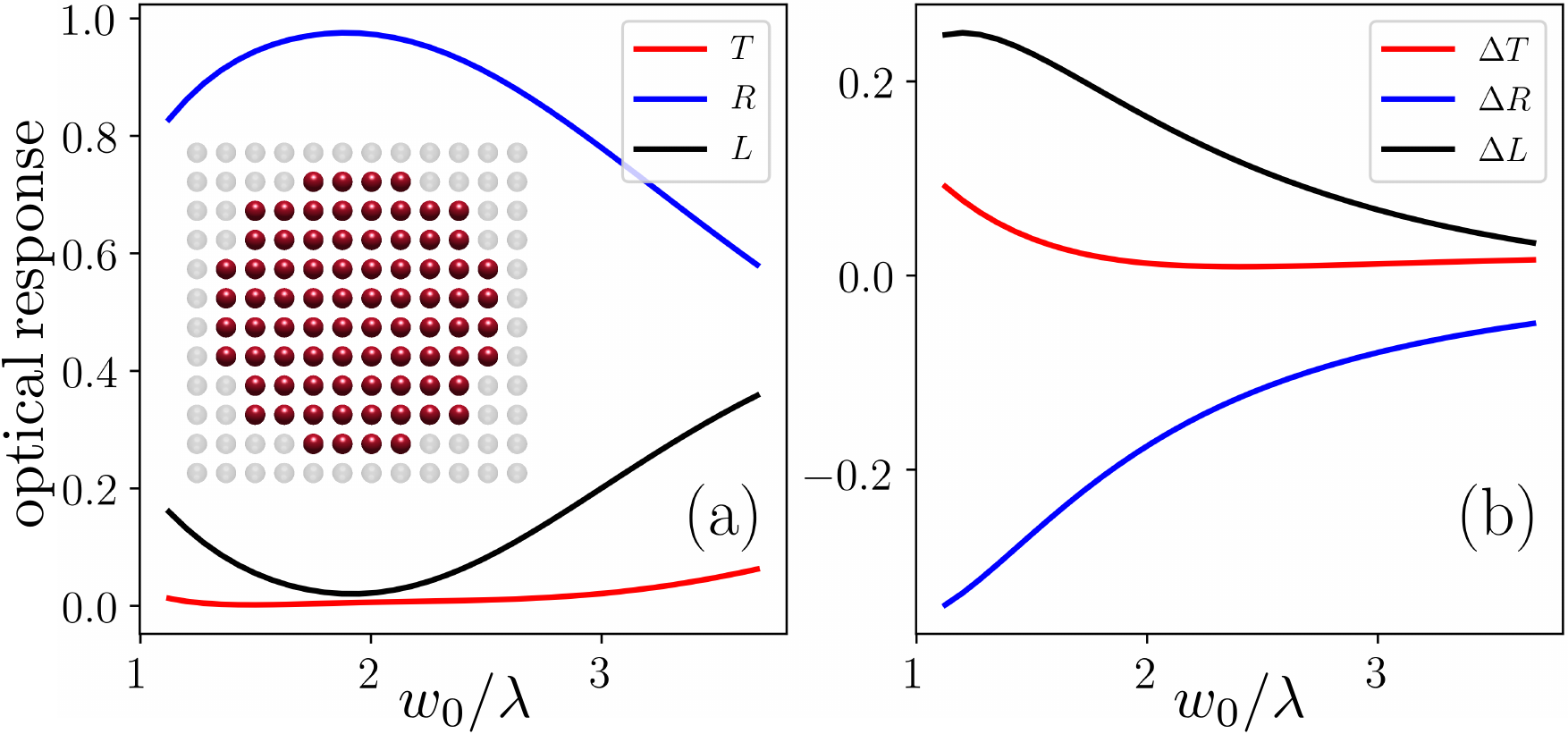}
 \caption{(a) Coefficients for linear reflection ($R$), transmission ($T$) and loss ($L$) of the incident probe light for an array of two-level atoms ($\Omega=0$) with a circular boundary as shown in the inset. The depicted dependence on the waist, $w_0$, of the probe beam shows a maximum, near-unity reflection of $R\simeq 0.975$ at $w_0\simeq 2\lambda$ for an optimized lattice constant $a=0.75\lambda$ and probe detuning $\Delta = 0.05\Gamma$. Panel (b) shows the average change of the linear response coefficients for identical parameters when adding a Rydberg defect in the form of an empty lattice site at ${\bf r}_j$ with a probability $p_j\propto |\mathcal{E}({\bf r}_j)|^2$. }
 \label{fig2}
\end{figure}

A quantum mechanical switching mechanism can emerge from the strong interaction between the Rydberg states. Hereby, the interaction blockade of multiple Rydberg excitation prevents the formation of more than a single dark-state excitation in the array, diminishing EIT for more than a single photon. Therefore, a single Rydberg excitation exposes the reflective two-level response to multi-photon states, while single probe photons can pass the array unimpeded. The Rydberg blockade, thus, switches between the two extreme cases of full transmission and full reflection, shown in Fig.~\ref{fig1}(e). As we shall see below, such a nonlinear switching mechanism may yield effective photon-photon interactions that can operate at the level of single photons and greatly suppressed scattering losses.
We have studied this behavior by solving the Master equation, $\partial_{t}\hat{\rho}(t)=-i[\hat{H},\hat {\rho}] + \mathcal{L}(\hat \rho)$, for the density matrix, $\hat{\rho}$, of the atomic array. The Hamiltonian, $\hat{H}=\hat{H}_{\rm LA}+\hat{H}_{\rm dd}$ contains the light-atom coupling in the rotating wave approximation
\begin{equation}\label{eq:Hal}
\hat{H}_{\rm LA}=-\sum^{N}_{j=1}\left[g\mathcal{E}({\bf r}_{j})\hat{\sigma}^{(j)}_{eg}
 + \Omega \hat{\sigma}^{(j)}_{es} + {\rm h.c.}\right]+\Delta\hat{\sigma}^{(j)}_{ee},
\end{equation}
where $\hat{\sigma}_{\alpha\beta}^{(j)}=|\alpha_j\rangle\langle \beta_j|$ denote the projection and transition operators for the $j$th atom, $g$ denotes the atom-photon coupling strength. The probe-field amplitude follows the paraxial wave equation $[4\pi i\partial_z+\lambda\nabla_\perp^2]\mathcal{E}=0$, and is normalized such that $|\mathcal{E}|^2$ yields a spatial photon density. The remaining photonic dynamics can be integrated out to obtain a Hamiltonian
\begin{equation}\label{eq:Hdd}
\hat{H}_{\text{dd}}=-\sum_{i\neq j}J_{ij}\hat{\sigma}^{(i)}_{eg}\hat{\sigma}^{(j)}_{ge}
\end{equation}
and Liouvillian
\begin{equation}\label{eq:L}
\mathcal{L}(\rho) = \sum^{N}_{i,j}\frac{1}{2}\Gamma_{ij}(2\hat{\sigma}^{(j)}_{ge}\rho\hat{\sigma}^{(i)}_{eg}-\hat{\sigma}^{(i)}_{eg}\hat{\sigma}^{(j)}_{ge}\rho-\rho\hat{\sigma}^{(i)}_{eg}\hat{\sigma}^{(j)}_{ge})
\end{equation}
\begin{figure}[t!]
 \centering
 \includegraphics[width=0.3\textwidth]{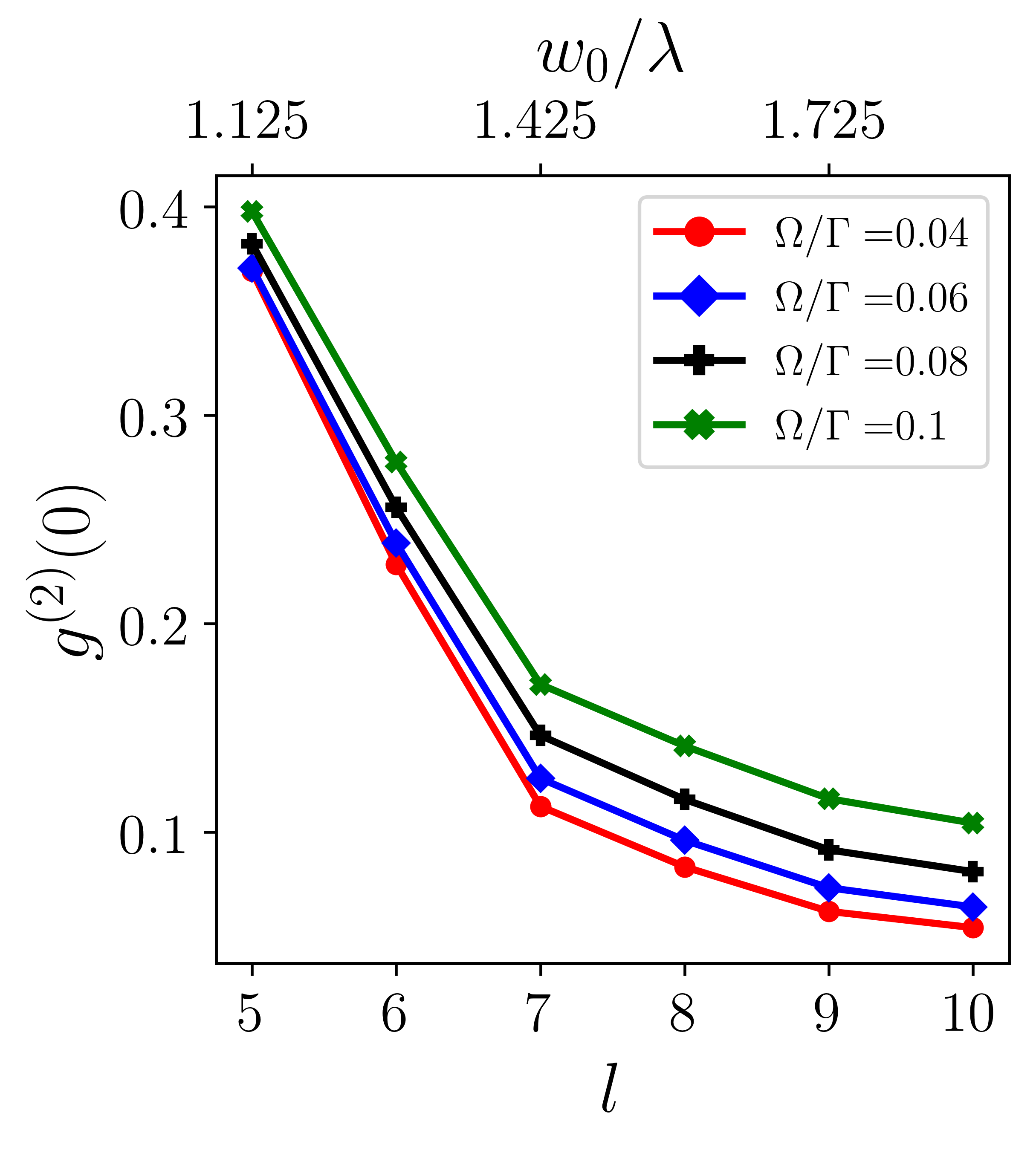}
 \caption{(a) Equal-time two-photon correlation function of the transmitted light as a function of diameter of the array for $\sqrt{\mathcal{P}} = 0.01\sqrt{\Gamma/c}$ and $a=0.75\lambda$. The beam waist is optimized to obtain a maximum linear reflection as shown in Fig.\ref{fig2}(a) and as given in the upper axis of the graph.}
 \label{fig3}
\end{figure}
that describe the light-induced atomic interactions within the Born-Markov approximation \cite{Garcia2017PRA}. The interaction coefficients $J_{ij}$ and $\Gamma_{ij}$ for two atoms at positions ${\bf r}_i$ and ${\bf r}_j$ are determined by the Greens function tensor of the free-space electromagnetic field. The transition dipole moment involving a high-lying Rydberg state and a low lying state is very small such that the corresponding dipole-dipole interactions can be safely neglected. For example, a tenfold increase of the Rydberg-state quantum number would give a suppression of such inter-state dipole-dipole interactions by 6 orders of magnitude. Knowing the dipolar field from each atom, one can readily reconstruct the mean values and correlation functions of the photonic field from the solution, $\hat{\rho}$, of the driven atomic many-body dynamics. While this yields the entirety of the emitted light field, we focus here on its projection onto the single transverse mode, $\mathcal{E}({\bf r})$, of the driving field. This yields simple relations
\begin{subequations}\label{eq:photons}
 \begin{align}
\label{eq:photons_a}  \hat{a}_{\rightarrow}(t)=&\hat{a}_{\rightarrow}^{({\rm in})}(t) + i\frac{g}{c\sqrt{\mathcal{P}}}\sum_{j}\mathcal{E}^{*}({\bf r}_j)\hat{\sigma}_{eg}^{(j)}(t),\\
\label{eq:photons_b}\hat{a}_{\leftarrow}(t)=&\hat{a}_{\leftarrow}^{({\rm in})}(t)  + i\frac{g}{c\sqrt{\mathcal{P}}}\sum_{j}\mathcal{E}^{*}({\bf r}_j)\hat{\sigma}_{eg}^{(j)}(t)
 \end{align}
\end{subequations}
for the photon operators of the transmitted ($\hat{a}_{\rightarrow}$) and reflected ($\hat{a}_{\leftarrow}$) field modes, in terms of the atomic transition operators. Here, $\mathcal{P}=\int|\mathcal{E}({\bf r})|^2{\rm d}{\bf r}_\perp$ denotes the transverse integral over the input intensity profile and defines the probe beam power, which is conserved along the propagation.
 Moreover, the operators $\hat{a}_{\rightarrow}^{({\rm in})}(t) $ and $\hat{a}_{\leftarrow}^{({\rm in})}(t)$ describe the field of the forward and backward propagating input mode, respectively. For the considered coherent drive in the forward direction, we have $\langle\hat{a}_{\rightarrow}^{({\rm in})}\rangle=\sqrt{\mathcal{P}}$ and $\langle\hat{a}_{\leftarrow}^{({\rm in})}\rangle=0$.

For reference, let us first consider the linear optical properties of the two-level array in the absence of Rydberg-state coupling ($\Omega=0$). While only an infinitely extended array can perfectly reflect an incident plane wave~\cite{Shahmoon2017PRL,Manzoni_2018}, finite arrays can yield high reflection for a judicious choice of the system parameters. This is illustrated in Fig.~\ref{fig2}(a), where we show the steady-state reflectivity $R=\langle\hat{a}_{\leftarrow}^\dagger\hat{a}_{\leftarrow}\rangle/\mathcal{P}$ along with the transmission coeffcient $T=\langle\hat{a}_{\rightarrow}^\dagger\hat{a}_{\rightarrow}\rangle/\mathcal{P}$ and loss $L=1-T-R$ for circular disc-shaped arrays with a diameter of $\ell$ atoms and a Gaussian driving mode $|\mathcal{E}|=\sqrt{2\mathcal{P}/(\pi w^2)}{\rm e}^{-r_\perp^2/w^2}$, whose width changes as $w^2=w_0^2+\lambda^2z^2/(\pi^2w_0^2)$ along the propagation direction and has its waist centered at the mirror position (Fig.\ref{fig1}). Already for an array with a $10$-atom diameter ($\ell=10$), one can reach near-unity reflection of $R\simeq 0.975$ for a beam waist of only $w_{0}\simeq 2\lambda$.

\begin{figure}[t!]
 \centering
 \includegraphics[width=0.48\textwidth]{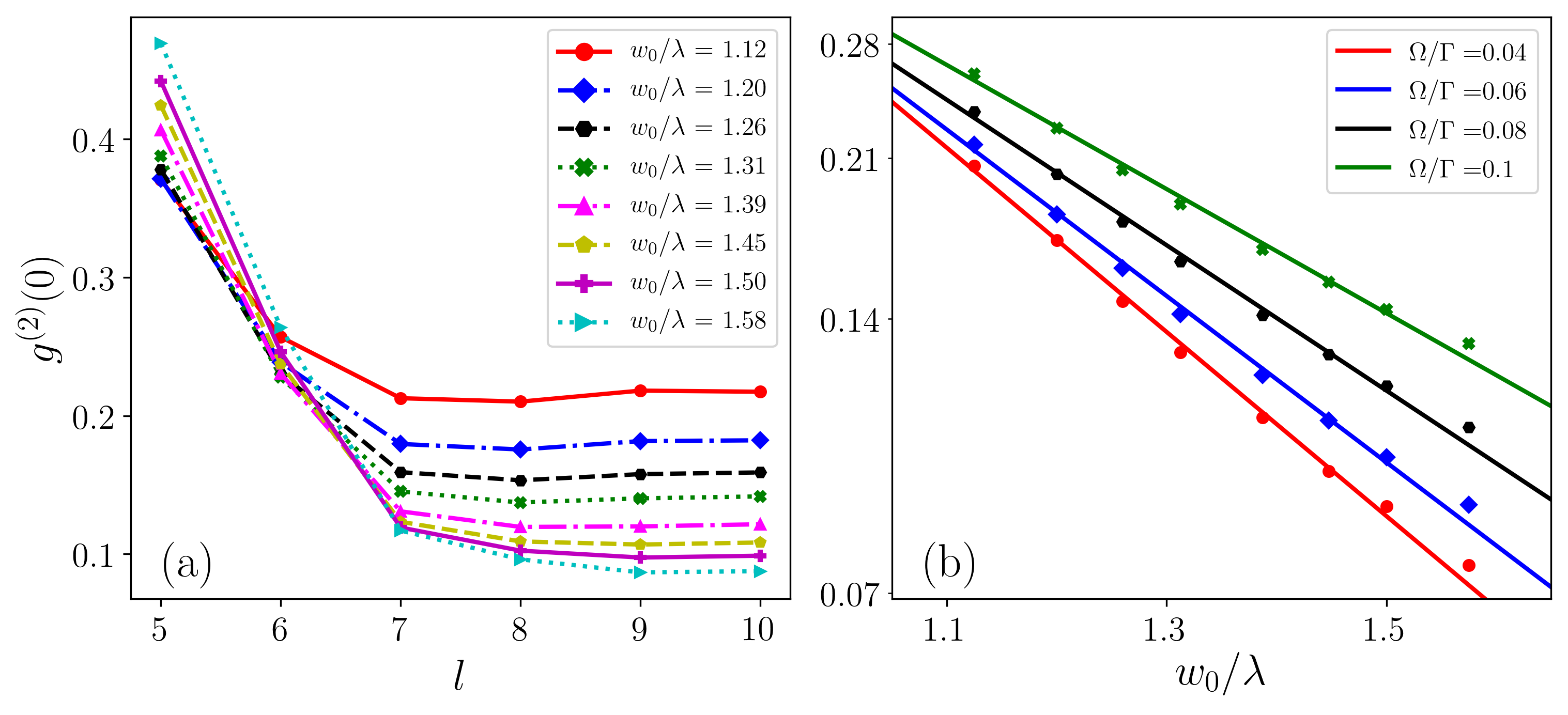}
 \caption{(a) Equal-time two-photon correlation function of the transmitted light as a function of diameter of the array for $\sqrt{\mathcal{P}} = 0.01\sqrt{\Gamma/c}$ and $a=0.75\lambda$. Results are shown for different values of the beam waist $w_0$, as indicated in the figure. From these results, we can extrapolate the asymptotic pair correlations for large arrays, $\ell\rightarrow\infty$ which are shown in panel (b) as a function of the probe-beam waist.}
 \label{fig4}
\end{figure}

The linear reflectivity for a finite Rydberg-state coupling, $\Omega$, is determined by the atomic dark state 
\begin{equation}\label{eq:darkstate}
|D\rangle\propto \Omega|G\rangle-g\sum_j\mathcal{E}({\bf r}_j)\hat{\sigma}_{sg}^{(j)}|G\rangle,
\end{equation}
where $|G\rangle$ denotes the $N$-atom ground state. Owing to the long Rydberg-atom lifetime, this state does not suffer from spontaneous emission and hence it facilitates a vanishing reflection and perfect transmission of the probe field. Under the typical condition $g\sqrt{\sum_j|\mathcal{E}({\bf r}_j)|^2}>\Omega$ a single incident photon is almost entirely converted into a Rydberg excitation \cite{Fleischhauer2000PRL} as it forms the Rydberg dark state, Eq.(\ref{eq:darkstate}). This switches the response from highly transmissive to highly reflective during this process. Subsequent photons that hit the array in the presence of this formed dark state excitation, are, thus, prevented from forming another dark state and reflected by the remaining two-level array. 
 
The photon-photon interaction that emerges from this mechanism is determined by the intricate interplay between EIT and the different atomic interactions, including (i) local defect interactions with the atomic Rydberg excitation, (ii) finite-range photon-mediated dipole-dipole interactions, and (iii) the long-range Rydberg interactions that extend across the array. 

The first effect simply refers to atomic saturation, preventing the generation of an $|e\rangle$-state excitation at the lattice site of the Rydberg atom. In addition to globally blocking any further Rydberg excitation, the Rydberg atom thus acts as a local defect for the low-lying excitations and thereby affects the reflectivity of the Rydberg-blockaded array. We can estimate the extent of this effect by sampling a single empty site of the array from the probability distribution $p_j\propto|\mathcal{E}({\bf r}_j)|^2$ of generated Rydberg impurities, according to the dark state Eq.(\ref{eq:darkstate}). As shown in Fig.\ref{fig2}(b), such a single de-localized Rydberg impurity can have observable consequences for the optical response, causing transverse photon scattering at the expense of the two-level reflection coefficient. Such losses, however, decrease with increasing beam waist as this decreases the defect density of the single Rydberg excitation according to $p_j\sim w_0^{-2}$.

In order to study the nonlinear response of the Rydberg array, we have performed stochastic wave function simulations \cite{Molmer93} to solve the $N$-body master equation determined by Eqs.(\ref{eq:Hal})-(\ref{eq:L}). For sufficiently weak probe fields, one can truncate the many-body wave function of the atoms at maximally two $|e\rangle$-excitations. This describes the physics of two interacting photons, which can be analyzed via the two-photon densities
\begin{equation}\label{eq:rho2}
\rho_{{\substack{\alpha\\ \beta}}}(t,t^\prime)=\langle\hat{a}_\alpha^\dagger(t)\hat{a}_\beta^\dagger(t^\prime)\hat{a}_\beta(t^\prime)\hat{a}_\alpha(t)\rangle ,
\end{equation}
and the associated correlation functions    $g^{(2)}_{{\substack{\alpha\\ \beta}}}(|t-t^\prime|) \equiv \rho_{{\substack{\alpha\\ \beta}}}(t,t^\prime) /(\langle\hat{a}_\alpha^\dagger(t)\hat{a}_\alpha(t)\rangle\langle\hat{a}_\beta^\dagger(t^\prime)\hat{a}_\beta(t^\prime)\rangle)$, where $\alpha,\beta=\rightarrow,\leftarrow$ label the forward and backward propagating mode of emitted probe photons, as also indicated in Eqs.(\ref{eq:photons}). The pair-correlation functions only depend on the time difference $\tau=|t-t^\prime|$ in the steady-state under cw-driving.

\begin{figure}[t!]
 \centering
 \includegraphics[width=0.48\textwidth]{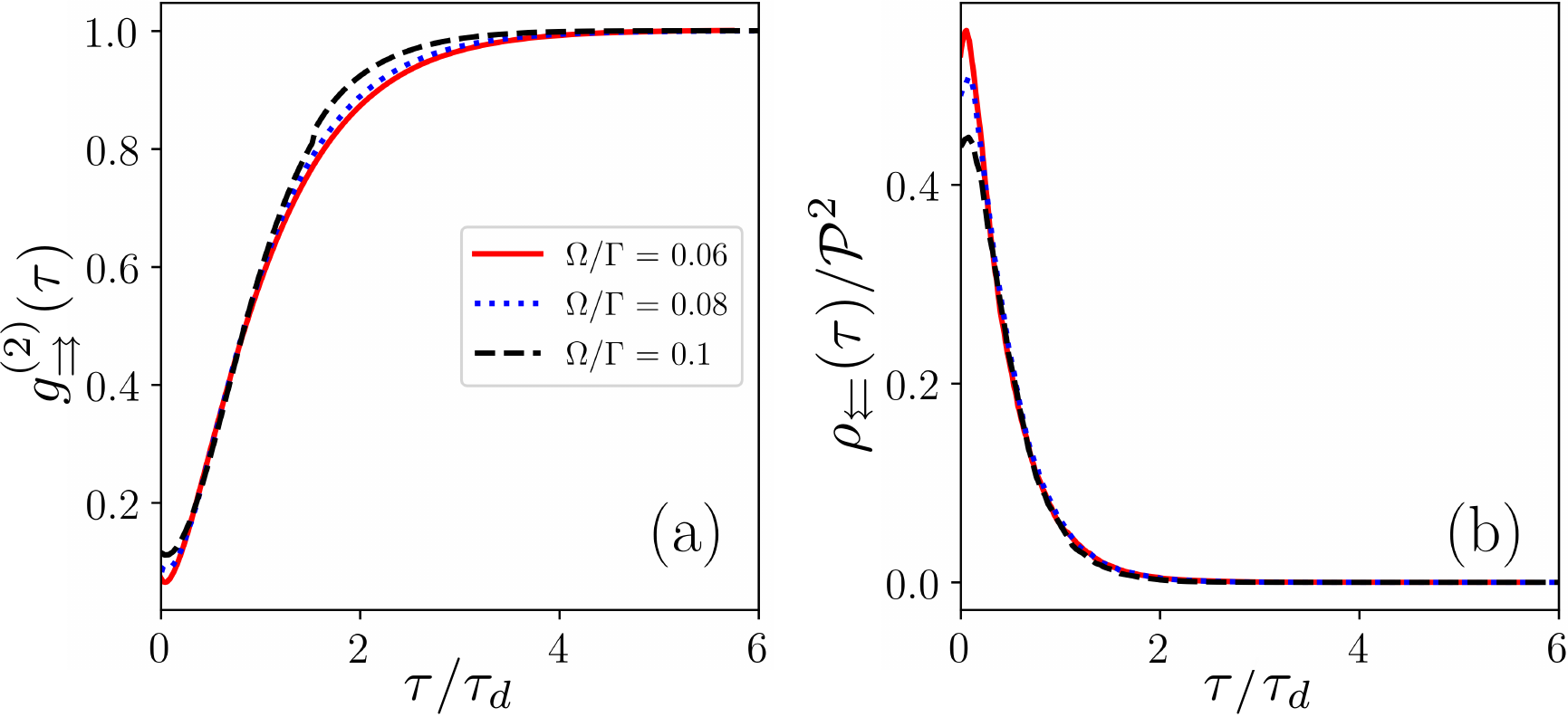}
 \caption{Pair correlation function of the transmitted (a) and reflected light (b) for different values of the control-field Rabi frequency $\Omega$, $\ell=10$, $w_{0}=1.7\lambda$, and otherwise identical parameters as in Fig.~\ref{fig3}. For both correlation functions, all data approximately collapse onto a single curve upon scaling the time between successive photon detections by the EIT delay time $\tau_d$, as given in Eq.(\ref{eq:td}).}
 \label{fig5}
\end{figure}

In Fig.\ref{fig3}, we show the equal-time two-photon correlations $g_{\rightrightarrows}^{(2)}(0)$ of the transmitted light for different sizes of the array, assuming parameters optimized to maximize the linear reflection. As can be seen, the Rydberg blockade can lead to strongly antibunched transmitted light. This is possible because the Rydberg-state component of the dark state, $|D\rangle$, generated by absorption of one photon, blocks excitation of further atoms into the dark state and therefore suppresses the simultaneous transmission of multiple photons. We find stronger anti-bunching for larger arrays, i.e. a rapid drop of $g_{\rightrightarrows}^{(2)}(0)$ with increasing size $\ell$ of the array. This behaviour arises from the effects of Rydberg impurities, discussed above and illustrated in Fig.\ref{fig2}(b). A larger size of the array and illuminated area implies a lower density of the single Rydberg impurity, and therefore improves the efficiency of the nonlinear reflection. In Fig.\ref{fig4}(a), we show simulations for a fixed beam waist and increasing lattice size, which demonstrates convergence to a finite value of $g_{\rightrightarrows}^{(2)}(0)$ for sufficiently large arrays. These asymptotic values are shown in Fig.\ref{fig4}(b) as a function of $w_0$. Over the numerically accessible range of beam waists and system sizes, we find a rapid exponential drop with increasing $w_0$ and strong anti-bunching with $g_{\rightrightarrows}^{(2)}(0)<0.1$ already for remarkably small values $w_0\sim1.5\lambda$.

These results demonstrate the strong suppression of multi-photon transmission, and the two-photon density depicted in Fig.\ref{fig1}(b) shows how incident photons are rerouted by their nonlinear interaction with the Rydberg array. Here, the two-time density defined in Eq.(\ref{eq:rho2}) has been converted to a spatial steady-state correlation function, using the relation between spatial photon position and time, $z=c t$ by the speed of light, $c$. In particular, we see that the two-photon density, $\rho_{\rightleftarrows}(z,z^\prime)=\rho_{\leftrightarrows}(z^\prime,z)$ for counter-propagating photon pairs continuously connects to the density, $\rho_{\leftleftarrows}(z,z^\prime)$, of simultaneously reflected photon pairs. This indicates that the two-photon component of the incident probe field is symmetrically rerouted into these two modes.

We can understand this behavior as follows. In the linear limit and in the absence of EIT ($\Omega=0$), the weak probe field $\mathcal{E}$ only drives weak atomic excitations with small transition dipole moments determined by $\hat{\sigma}_{eg}^{(j)}$. At the reflection maxima, the field generated by these weak atomic dipoles just cancels the probe field in Eq.(\ref{eq:photons_a}) and therefore yields high reflection with $\langle\hat{a}_{\leftarrow}\rangle\approx\sqrt{\mathcal{P}}$ according to Eq.(\ref{eq:photons_b}). In the opposite limit of perfect EIT, the atomic dipoles vanish entirely, leading to perfect transmission with $\langle\hat{a}_{\rightarrow}\rangle=\sqrt{\mathcal{P}}$ and $\langle\hat{a}_{\leftarrow}\rangle=0$, according to Eqs.(\ref{eq:photons}). The nonlinear response, however, differs fundamentally, because the detection of the first reflected photon causes a projection of the $N$-atom wavefunction into a state with a definite de-localized excitation. This unit-probability, heralded excitation consequently contributes a much stronger emission that vastly overwhelms the incident probe fields, $\hat{a}_{\rightarrow}^{({\rm in})}$ and $\hat{a}_{\leftarrow}^{({\rm in})}$. Following Eq.(\ref{eq:photons}), the subsequent conditioned photon emission, therefore, becomes virtually symmetric, with $\hat{a}_{\rightarrow}\approx\hat{a}_{\leftarrow}$ and leads to the typical form of the correlated two-photon density shown in Fig.\ref{fig1}(b).

Fig.4 offers further insights into the dynamics of the nonlinear photon interaction. From the linear response of the array under EIT conditions, we find that a transmitted light pulse experiences a delay of
\begin{equation}\label{eq:td}
\tau_d=\frac{\Gamma_c}{2\Omega^2},
\end{equation}
which coincides with the inverse width of the transparency window shown in Fig.\ref{fig1}(e). This pulse delay emerges in analogy to slow-light propagation through an extended EIT medium \cite{Fleischhauer2000PRL}, and corresponds to the average time for which a transmitted photon is transferred to the de-localized Rydberg state $\sim \sum_j\mathcal{E}({\bf r}_j)\hat{\sigma}_{sg}^{(j)}|G\rangle$ and blocks EIT for any other incident photons. The pair correlation functions and two-photon densities, depicted in Fig.\ref{fig4}, accurately corroborate this picture, showing the same characteristic correlation time $\tau_d$ for bunched and anti-bunched photon states of the reflected and transmitted light for varying values of the control-field Rabi frequency $\Omega$. At the same time, we find that the outgoing photons maintain a high degree of coherence, as quantified by $g_{\alpha}^{(1)}(t)=\langle\hat{a}_\alpha^\dagger(t)\hat{a}_\alpha(0)\rangle/\langle\hat{a}_\alpha^\dagger(0)\hat{a}_\alpha(0)\rangle\sim1$, on the scale of the characteristic correlation time of both fields, which reflects the suppression of photon loss and decoherence by the collective light-matter coupling of the ordered array. 

This combination of high coherence, low photon losses and strong photon-photon interactions offers a promising outlook for the generation and manipulation of non-classical light in optical-lattice experiments that have already  demonstrated Rydberg blockade of more than $\sim100$ atoms \cite{Zeiher2015} as well as efficient photon reflection by arrays with sub-wavelength lattice constants \cite{Rui2020Nature}. The demonstrated nonlinearity is akin to that of wave guide QED with single few-level emitters, whereby the eliminated scattering into other transverse modes effectively corresponds to a near-perfect coupling into a single guided mode. This limit of strong coherent photon coupling has thus far been difficult to reach in atomic systems \cite{Prasad2020,Stiesdal2021}, but will enable a range of applications, from the generation of single narrow-bandwidth photons \cite{Parkins1993}, and logic gates \cite{Ralph2015} to few-photon routing and sorting \cite{Witthaut2012}. The Rydberg array can hereby be employed as an active or passive element under pulsed or cw operation, exploiting the additional temporal control provided by the control-field coupling. 
While we have focussed here on the few-photon domain in order to analyse the basic interaction mechanism, the multi-photon regime under strong-driving conditions provides an exciting perspectives for exploring quantum optical many-body phenomena. 
Similarly to cavity-QED with single emitters, the described nonlinearities may be further enhanced by positioning the array in front of mirrors or inside optical cavities \cite{shahmoon2020}. Arrangements of multiple Rydberg arrays, or more complex 3D configurations could be constructed with atoms in configurable optical tweezer arrays to form networks of quantum beam splitters and nonlinear resonators that also exploit multi-photon and multi-mode interference effects and may supplement proposals for quantum enhanced interferometry \cite{Demkowicz2015}.\\

This work was supported by the Carlsberg Foundation through the 'Semper Ardens' Research Project QCooL, by the NSF through a grant for ITAMP at Harvard University, by the DFG through the SPP1929, by the European Commission through the H2020-FETOPEN project ErBeStA (No. 800942), and by the Danish National Research Foundation through the Center of Excellence "CCQ" (Grant agreement no.: DNRF156).\\

\emph{Note added}: During completing of this manuscript we became aware of a related work \cite{Moreno2021} that describes nonlinearities in two-level arrays where finite-range interactions are induced by Rydberg dressing \cite{henkel2010}.

%

%\bibliographystyle{apsrev4-1}
%\bibliographystyle{apsrev4-1}
%\bibliography{references}
\end{document}